\documentclass[aps,onecolumn]{revtex4-2}

\usepackage[T1]{fontenc}
\usepackage{charter}

\usepackage{amsmath,amssymb,amsfonts,color,graphicx,tabularx}

\usepackage[unicode=true,colorlinks=true]{hyperref}

\hypersetup{linkcolor=blue,citecolor=blue,urlcolor=blue}

\begin{document}

\title{Third-order quantum phase transitions of bosonic \\ non-Abelian fractional quantum Hall states}

\author{Kai-Wen Huang}
\author{Xiang-Jian Hou}
\affiliation{School of Physics and Wuhan National High Magnetic Field Center, Huazhong University of Science and Technology, Wuhan 430074, China}

\author{Ying-Hai Wu}
\email{yinghaiwu88@hust.edu.cn}
\affiliation{School of Physics and Wuhan National High Magnetic Field Center, Huazhong University of Science and Technology, Wuhan 430074, China}

\begin{abstract}
We study phase transitions in bilayer and trilayer bosonic quantum Hall systems. In the absence of interlayer tunneling and interaction, each layer is chosen to have filling factor $\nu=1/2$ or $1$ to realize the Laughlin state or the Moore-Read state. By tuning interlayer tunneling and/or interaction, multiple phases can be generated. In the absence of interlayer interaction, three phase transitions appear when interlayer tunneling becomes sufficiently strong: (1) from two decoupled $\nu=1/2$ Laughlin states to the Moore-Read state in bilayer systems; (2) from one $\nu=1/2$ Laughlin state plus one $\nu=1$ Moore-Read state to the Read-Rezayi $\mathbb{Z}_{3}$ state in bilayer systems; (3) from three decoupled $\nu=1/2$ Laughlin states to the Read-Rezayi $\mathbb{Z}_{3}$ state in trilayer systems. Numerical calculations suggest that these transitions are third-order ones. We propose non-Abelian Chern-Simons-Higgs theory to describe them. If  both interlayer tunneling and interaction are present, one-component or multi-component composite fermion liquids and Jain states can be realized. This leads to intricate phase diagrams that host multiple phase transitions and possibly exotic critical points.
\end{abstract}

\maketitle

\section{Introduction}
\label{intro}

The quantum mechanical motion of an electron in a magnetic field was solved by Landau~\cite{Landau1930}. If the motion is frozen along one direction but unconstrained in a plane perpendicular to it, the single-particle eigenstates would form highly degenerate Landau levels (LLs) in which each orbital is associated with one magnetic flux $h/e$. While this is a very interesting feature, its significance only became apparent with the discovery of quantum Hall effect~\cite{Klitzing1980,TsuiDC1982}. If electron-electron interactions are neglected, a partially filled LL has many different configurations with equal kinetic energy. This turns out to be an excellent setting for probing strongly correlated phenomena because many competing states have nearly the same energy. Fractional quantum Hall (FQH) effect is arguably the most prominent example. It serves as a conspicuous illustration of non-perturbative emergence of exotic properties in a macroscopic collection of simple objects~\cite{Halperin-Book}. An intriguing property of FQH states is that some elementary excitations are anyons~\cite{Laughlin1983,Leinaas1977,Halperin1984,Arovas1984}. It is intimated related to another fact: a FQH system has multiple degenerate ground states when it is placed on surfaces with nonzero genus such as the torus. This fact motivated the concept of topological order for which FQH states are canonical examples~\cite{WenXG-Book}. 

Going beyond the most common scenario, FQH states have also been explored in other settings. In theoretical studies, it is simple to replace electrons with bosons, and FQH states can still be realized provided appropriate interactions are used~\cite{Cooper2008,Viefers2008}. There is no negatively charged boson at our dispose, but effective LLs can be synthesized for atomic gases and photons to realize small FQH droplets~\cite{Gemelke2010,Clark2019,Fletcher2021,Mukherjee2022}. More generally, FQH physics is not restricted to LLs. From the band theory perspective, Hall response is determined by the Chern number of bands~\cite{Thouless1982}, and nonzero value can appear in lattice models without magnetic field~\cite{Haldane1988b}. Fractional Chern insulators may be found in an isolated Chern band with suitable conditions~\cite{LiuZ2024}, which are similar to FQH states in many aspects but also exhibit differences. Experimental demonstrations of this route have been reported in van der Waals heterostructures~\cite{CaiJQ2023,ZengYH2023,ParkHJ2023,XuF2023,LuZG2024} and artificial quantum systems~\cite{Leonard2022,LiuFM2024}.

For a given set of microscopic ingredients, one main task is to elucidate the possible phases and transitions between them. The Landau-Ginzburg-Wilson framework has been very successful in dealing with symmetry-breaking phases~\cite{Sachdev-Book}, but it is not applicable in FQH states that lack order parameters. In this work, we investigate phase transitions in three multi-layer bosonic systems. A plethora of states can be realized using different combinations of intralayer interaction, interlayer interaction, and interlayer tunneling, including the Laughlin state~\cite{Laughlin1983}, the Jain states~\cite{Jain1989a}, the Moore-Read state~\cite{Moore1991}, the Read-Rezayi $\mathbb{Z}_{3}$ state~\cite{Read1999}, and the composite fermion liquid~\cite{Halperin1993}. When the interlayer tunneling is varied with other factors chosen properly, three third-order phase transitions can be realized. Ehrenfest proposed to define a thermal phase transition as $n$-th order if the free energy has a discontinuity in its $n$-th derivative~\cite{Goldenfeld-Book}. It was later realized that certain derivatives actually diverge. The classification may be modified such that there are only first-order ones and continuous ones. For quantum phase transitions, the free energy could be replaced by the ground state energy. In our systems, third-order derivatives of the ground state energy exhibit divergences. This kind of behavior is quite rare. Some examples were found in Bose-Einstein condensation of free bosons, $p$ wave superconductors~\cite{Rombouts2010}, and Chern insulators~\cite{YouYZ2013}. 

\section{Models and States}
\label{model}

The Landau problem can be solved on different surfaces. When discussing many-body wave functions, it is convenient to employ the infinite disk with symmetric gauge. For the lowest LL, the single-particle eigenstates can be simplified to $\sim z^{m}$ ($m\in\mathbb{N}$) by dropping other inconsequential factors. This means that a many-body wave function in the lowest LL is a multivariate polynomial of all electron coordinates written as complex numbers. For numerical calculations, compact geometries such as sphere and torus~\cite{Haldane1983c,YoshiokaD1983} are more useful because they are free of the complications caused by edges. In both cases, the number of magnetic fluxes through the surface is an integer. For the sphere, the single-particle eigenstates are labeled by their angular momentum along the $z$ axis. For the torus, the single-particle eigenstates are labeled by a momentum along the $y$ axis. In our subsequent discussions, the aspect ratio of the torus is fixed at $1$. We study bosons in the lowest LL and express their interactions using the Haldane pseudopotentials~\cite{Haldane1983c}. In second quantized notation, the many-body Hamiltonian is written as
\begin{eqnarray}
\frac{1}{2} \sum_{\{m_{i}\}} V_{m_{1}m_{2}m_{3}m_{4}} C^{\dag}_{m_{1}} C^{\dag}_{m_{2}} C_{m_{3}} C_{m_{4}}.
\label{ManyBodyHami}
\end{eqnarray}
Here we have assumed that the particles do not possess any internal degrees of freedom. Angular momentum on the sphere or momentum on the torus is denoted by the subscripts $m_{i}$ and $C^{\dag}_{m}$ ($C^{\dag}_{m}$) is the creation (annihilation) operator associated with the orbital labeled by $m$. For each pseudopotential, the coefficients $V_{m_{1}m_{2}m_{3}m_{4}}$ can be evaluated using standard methods~\cite{Jain-Book}. 

Next we turn to trial wave functions for one-component bosons (i.e. there is no internal degree of freedom). The number of bosons is denoted as $N_{b}$. At filling factor $\nu=1/2$, we have the Laughlin wave function $\prod^{N_{b}}_{j<k}(z_{j}-z_{k})^{2}$~\cite{Laughlin1983}. It is actually the highest density zero-energy eigenstate of the zeroth Haldane pseudopotential. Non-Abelian Moore-Read and Read-Rezayi $\mathbb{Z}_{k}$ states are constructed from multi-point correlation functions of the SU(2)$_{k}$ conformal field theory~\cite{Moore1991,Read1999}. This approach provides a clear illustration of bulk-boundary correspondence and elucidates many properties about non-Abelian anyons. To write down their wave functions, the bosons are divided equally into $k$ groups with coordinates $\{w^{(a)}_{\alpha}\}$ ($a=1,2,\cdots,k$). Each group realizes a Laughlin state that are multiplied together and symmetrized with respect to the division~\cite{Cappelli2001,Repellin2015}. The final result is
\begin{eqnarray}
\Psi_{k} (\{z\}) = \mathcal{S} \left[ \prod^{N_{b}/k}_{\alpha<\beta} (w^{(1)}_{\alpha}-w^{(1)}_{\beta})^{2} \prod^{N_{b}/k}_{\alpha<\beta} (w^{(2)}_{\alpha}-w^{(2)}_{\beta})^{2} \cdots \prod^{N_{b}/k}_{\alpha<\beta} (w^{(k)}_{\alpha}-w^{(k)}_{\beta})^{2} \right].
\label{NonAbelianWave}
\end{eqnarray}
and it is the highest density zero-energy eigenstate of a $(k+1)$-body contact potential. If $\Psi_{k}(\{z\})$ is defined on the sphere, the number of fluxes is $N^{\rm S}_{\phi}=2N_{b}/k-2$. If $\Psi_{k}(\{z\})$ is defined on the torus, the number of fluxes is $N^{\rm T}_{\phi}=2N_{b}/k$ and the ground state degeneracy is $k+1$. We only consider the cases with $k=2$ and $3$. The former one is the Moore-Read wave function for which the symmetrization can be done explicitly to yield a Pfaffian. The latter one is the Read-Rezayi $\mathbb{Z}_{3}$ wave function that has no simple expression. Previous numerical calculations have confirmed the existence of these states~\cite{Cooper2001,Regnault2004a,ChangCC2005,Rezayi2005,Regnault2007,Cooper2007}.

\begin{figure}[ht]
\centering
\includegraphics[width=0.60\textwidth]{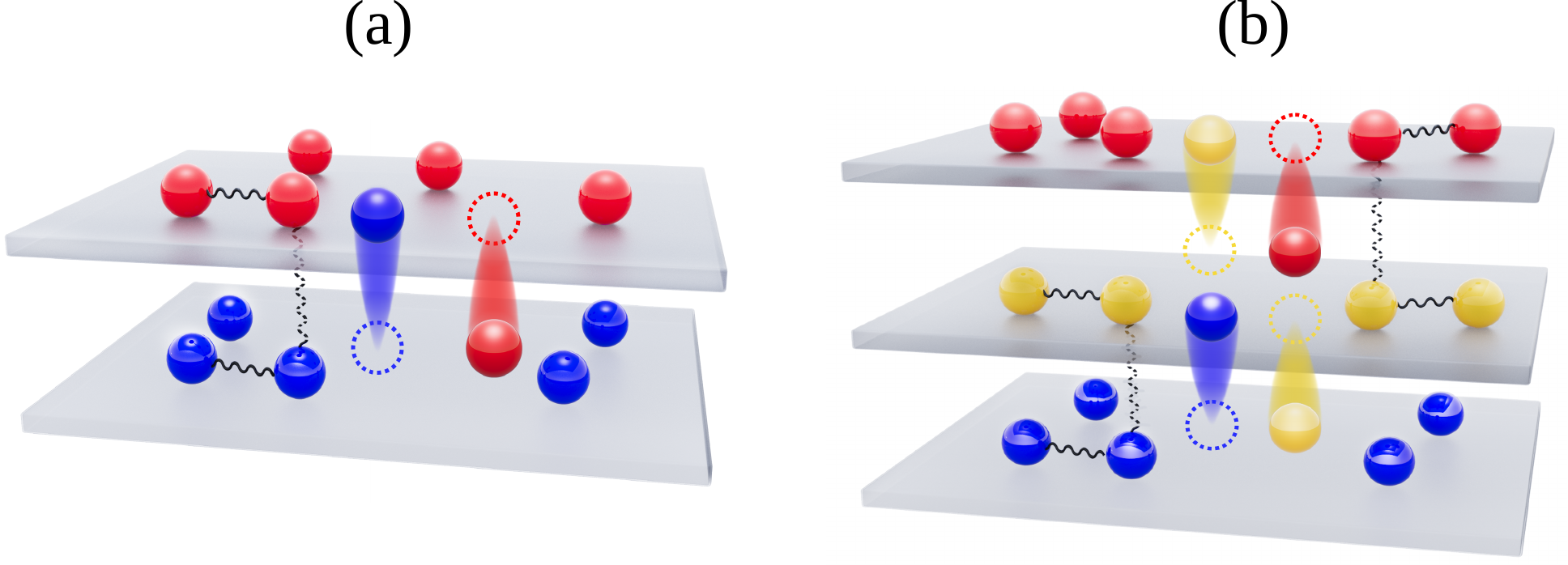}
\caption{Schematics of the bilayer and trilayer systems. In both cases, there are intralayer interactions (solid curves) and tunneling between adjacent layers (shaded arrows). We do not need interaction interactions (dashed curves) for the transitions studied in Sec.~\ref{third} but they are introduced in Sec.~\ref{other}.}
\label{Figure1}
\end{figure}

While these states are defined for one-component bosons, we shall study multi-layer systems as illustrated in Fig.~\ref{Figure1}. If there are two layers, they are called $\mathsf{t}$ and $\mathsf{b}$. If there are three layers, they are called $\mathsf{t}$, $\mathsf{m}$, and $\mathsf{b}$. These symbols are used in various places to label physical quantities. For example, the filling factors in bilayer systems are denoted as $\nu_{\mathsf{t}}$ and $\nu_{\mathsf{b}}$. The creation (annihilation) operators for the top layer is denoted as $C^{\dag}_{m}(\mathsf{t})$ [$C_{m}(\mathsf{t})$] and similar for the other layers. Since the bosons are allowed to tunnel between different layers, the number of bosons in each layer is not conserved. This means that we still have essentially a one-component system but there are important differences. For the bilayer case, the two sets of LLs can be reorganized as symmetric and antisymmetric bands in the presence of tunneling. The splitting between them is proportional to the tunneling strength. It is essential to keep both bands in our study. One can similarly define more complicated combinations in the trilayer case. Roughly speaking, interlayer tunneling provides a physical implementation of symmetrization in Eq.~\eqref{NonAbelianWave}. It has been demonstrated that bosonic or fermionic Moore-Read states indeed arise in certain bilayer systems with tunneling~\cite{ZhuW2015a,LiuZ2016a,Crepel2019b,Cabra2001,Peterson2010a,Papic2010a,Peterson2010b,ZhuW2016c}.

\section{Third-Order Phase Transitions}
\label{third}

This section presents our main results about three third-order phase transitions. The first two are in bilayer systems and the last one is in trilayer systems.

\subsection{Bilayer systems with $\nu_{\mathsf{t}}=1/2$ and $\nu_{\mathsf{b}}=1/2$}

In this example, the system realizes two decoupled Laughlin states in the absence of interlayer tunneling and turns into a Moore-Read state when the tunneling becomes sufficiently large. This case has been studied in a few previous works~\cite{ZhuW2015a,LiuZ2016a,Crepel2019b} but the transition itself was not examined in detail. We study the many-body Hamiltonian
\begin{eqnarray}
H_{1} = V_{0}({\mathsf{t}}) + V_{0}(\mathsf{b}) - t \sum_{k} \left[ C^{\dag}_{k}(\mathsf{t}) C_{k}(\mathsf{b}) + C^{\dag}_{k}(\mathsf{b}) C_{k}(\mathsf{t}) \right].
\end{eqnarray}
Bosons in the same layer interact via the zeroth Haldane pseudopotentials as reflected by the first and second terms. There is no interlayer interaction for now. The third term represents interlayer tunneling that conserves angular momentum on the sphere or momentum on the torus. In terms of second quantized operators, $V_{0}(\mathsf{t})$ is similar to Eq.~\eqref{ManyBodyHami} except that $\mathsf{t}$ is included here to represent the layer degree of freedom.

\begin{figure}[ht]
\centering
\includegraphics[width=0.95\textwidth]{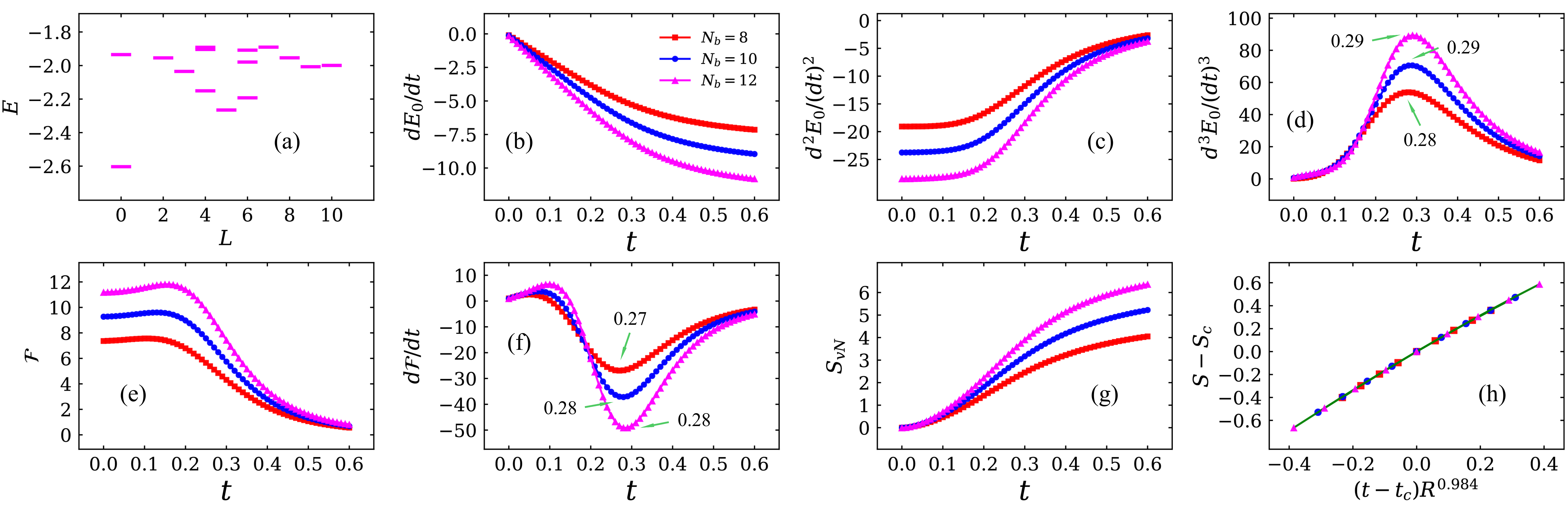}
\caption{Numerical results about bilayer systems with $\nu_{\mathsf{t}}=1/2,\nu_{\mathsf{b}}=1/2$ on the sphere. (a) Energy spectrum for the case with $N_{b}=12$ and $t=-0.45$. (c-d) First-, second-, and third-order derivatives of the ground state energy. (e-f) Fidelity susceptibility of the ground state and its first-order derivative. (g-h) Entanglement entropy between the two layers and its data collapse in the vicinity of $t_{c}=0.29$. The numbers in panels (d) and (f) are horizontal coordinates of the peaks.}
\label{FigureS1}
\end{figure}

Numerical results on the sphere are presented in Fig.~\ref{FigureS1}. We plot eigenvalues versus the total angular momentum $L$ for $N_{b}=12$ and $t=-0.45$ in panel (a). The system has entered the Moore-Read phase (see below). In addition to the ground state, a branch of neural excitations at $L=2,3,4,5,6$ is identified. For the Laughlin state, neutral excitations can be constructed as composite fermion excitons~\cite{ChangCC2005}. If there are six bosons, their angular momenta are $2,3,4,5,6$. Neutral excitations of the Moore-Read state can be obtained from those of the Laughlin state~\cite{Sreejith2011a,Rodriguez2012a}. Because of the symmetrization factor in Eq.~\eqref{ManyBodyHami}, the two sets of neutral excitations associated with individual Laughlin states are in fact equivalent. This explains why there is only one branch in panel (a). As one can see from panels (b-d), the first- and second-order derivatives of the ground state energy $E_{0}$ are continuous, but the third-order derivatives have peaks whose heights increase with $N_{b}$. The positions of maximum on each curve are indicated in panel (d). We conclude that a third-order phase transition occurs at $t_{c} \approx 0.29$. It has been established in previous works that fidelity susceptibility of the ground state is an excellent probe of phase transitions~\cite{YouWL2007,Cozzini2007}. For two adjacent parameters $t$ and $t+\delta{t}$, the corresponding ground states are denoted as $\Psi_{0}(t)$ and $\Psi_{0}(t+\delta{t})$, and the fidelity susceptibility is
\begin{eqnarray}
\mathcal{F}(t) = \frac{2}{(\delta t)^{2}} \left[ 1 - \left| \langle \Psi(t) | \Psi(t+\delta{t})\rangle \right| \right].
\end{eqnarray}
This quantity is displayed in panel (e) but there is no obvious peak. After taking the derivative with respect to $t$, peaks appear at positions that are close to $t_{c}$ as shown in panel (f). We speculate that fidelity susceptibility may not be able to diagnose higher-order phase transitions, but this requires more in-depth examinations. In the vicinity of a critical point, many quantities such as the order parameter follow scaling laws. Since there is no order parameter in our system, we turn to the von Neumann entanglement entropy $S_{\rm vN}(t)$ between the two layers. Its value in the full range of $t$ is given in panel (g) and data collapse for $t\in[0.24,0.33]$ is shown in panel (h). The scaling function is 
\begin{eqnarray}
S_{\rm vN}(t)-S_{{\rm vN}}(t_{c})=f\left[(t-t_{c})R^{\alpha}\right],
\label{EntropyScaleFunc}
\end{eqnarray}
where $S_{{\rm vN},c}$ is the entropy at the critical point, $R=\sqrt{N^{\rm S}_{\phi}}$ is radius of the sphere, and $\alpha=0.984$.

\begin{figure}[ht]
\centering
\includegraphics[width=0.95\textwidth]{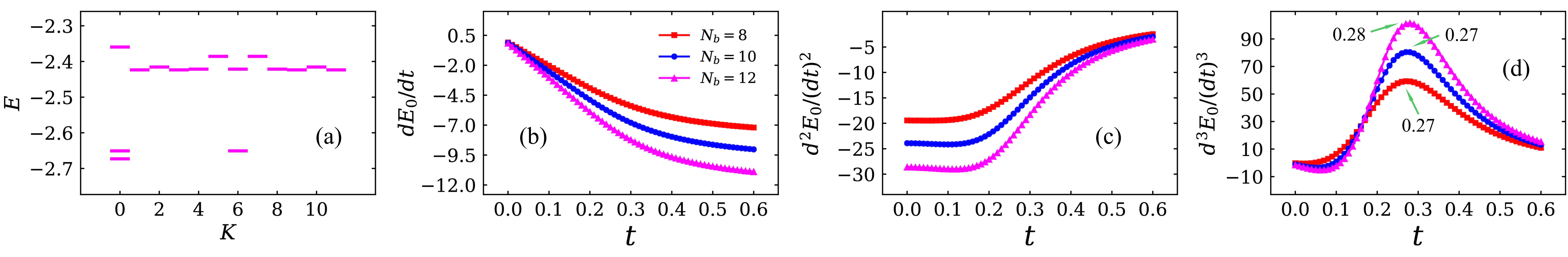}
\caption{Numerical results about bilayer systems with $\nu_{\mathsf{t}}=1/2,\nu_{\mathsf{b}}=1/2$ on the torus. (a) Energy spectrum for the case with $N_{b}=12$ and $t=-0.45$. (b-d) First-, second-, and third-order derivatives of the ground state energy. The numbers in panel (d) are horizontal coordinates of the peaks.}
\label{FigureT1}
\end{figure}

Numerical results on the torus are presented in Fig.~\ref{FigureT1}. Two decoupled Laughlin states on the torus have ground state degeneracy four whereas the Moore-Read state has three. As shown in panel (a), three quasi-degenerate ground states appear in the case with $N_{b}=12$ and $t=-0.45$. The horiztonal axis in this plot is the total momentum $K$ along the $y$ axis of the torus. Due to this quasi-degeneracy, the term ``ground state energy" is somewhat ambiguous. For our purpose, it is defined as the average of the lowest three eigenvalues. Its derivatives are displayed in panels (b-d). The first- and second-order derivatives are continuous but the third-order derivatives have peaks whose heights increase with $N_{b}$. This suggests that a phase transition occurs at $t_{c} \approx 0.28$. It is propitious that the critical points in two different geometries are very close.

\subsection{Bilayer systems with $\nu_{\mathsf{t}}=1/2$ and $\nu_{\mathsf{b}}=1$}

In this example, the system realizes decoupled Laughlin and Moore-Read states in the absence of interlayer tunneling and turns into a Read-Rezayi $\mathbb{Z}_{3}$ state when the tunneling becomes sufficiently large. We study the many-body Hamiltonian
\begin{eqnarray}
H_{2} = V_{0}(\mathsf{t}) + 0.2 V_{2}(\mathsf{t}) + V_{0}(\mathsf{b}) + 0.2 V_{2}(\mathsf{b}) + 1.5 \sum_{k} C^{\dag}_{k}(\mathsf{t}) C_{k}(\mathsf{t}) - t \sum_{k} \left[ C^{\dag}_{k}(\mathsf{t}) C_{k}(\mathsf{b}) + C^{\dag}_{k}(\mathsf{b}) C_{k}(\mathsf{t}) \right].
\end{eqnarray}
For each layer, both the zeroth and the second Haldane pseudopotentials are used. This combination is employed to stabilize the Moore-Read state in finite-size systems. It is known that the zeroth pseudopotential alone can realize the Moore-Read state~\cite{Cooper2001,Regnault2004a,ChangCC2005}, but the ground state degeneracy on the torus is not good when the number of bosons is relatively small. The fifth term is an additional potential for the top layer. It ensures that, in the absence of interlayer tunneling, we do have decoupled Laughlin and Moore-Read states. If there is no such term, the lowest energy configuration at $t=0$ actually appears at $\nu_{\mathsf{t}}=3/4,\nu_{\mathsf{b}}=3/4$, and it is not possible to study the desired phase transition.

\begin{figure}[ht]
\centering
\includegraphics[width=0.95\textwidth]{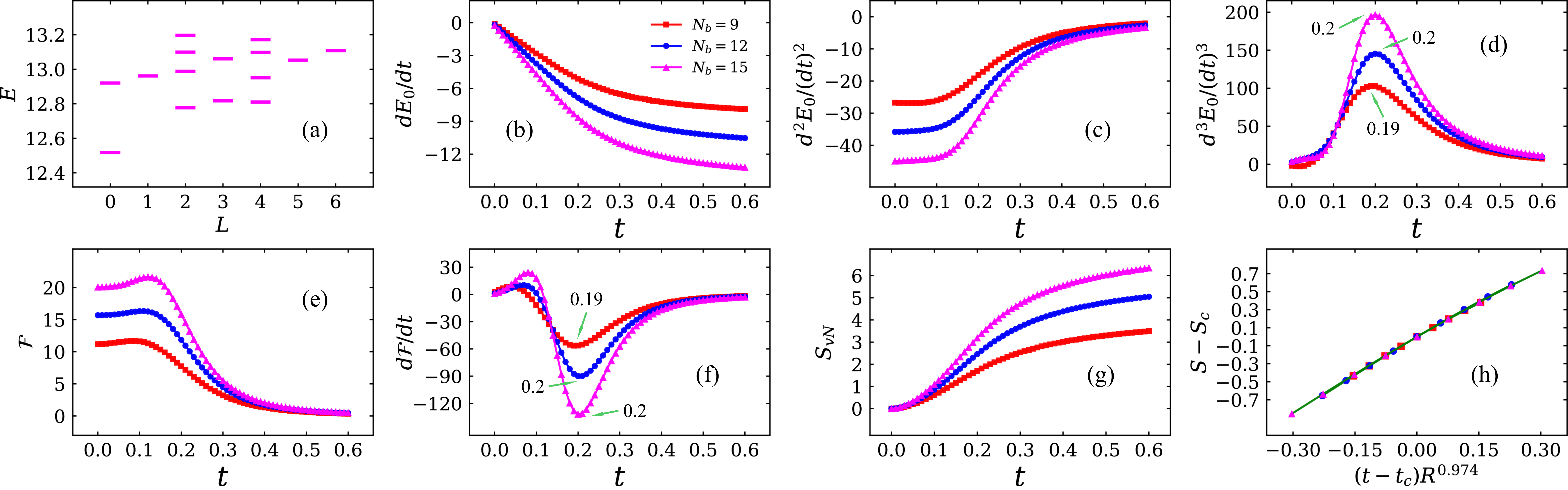}
\caption{Numerical results about bilayer systems with $\nu_{\mathsf{t}}=1/2,\nu_{\mathsf{b}}=1$ on the sphere. (a) Energy spectrum for the case with $N_{b}=12$ and $t=-0.45$. (b-d) First-, second-, and third-order derivatives of the ground state energy. (e-f) Fidelity susceptibility of the ground state and its first-order derivative. (g-h) Entanglement entropy between the two layers and its data collapse in the vicinity of $t_{c}=0.20$. The numbers in panels (d) and (f) are horizontal coordinates of the peaks.}
\label{FigureS2}
\end{figure}

\begin{figure}[ht]
\centering
\includegraphics[width=0.95\textwidth]{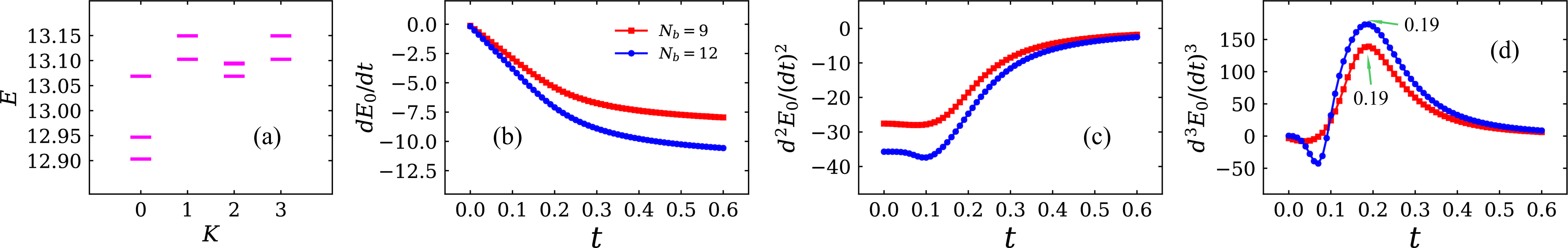}
\caption{Numerical results about bilayer systems with $\nu_{\mathsf{t}}=1/2,\nu_{\mathsf{b}}=1$ on the torus. (a) Energy spectrum for the case with $N_{b}=12$ and $t=-0.45$. (b-d) First-, second-, and third-order derivatives of the ground state energy. The numbers in panel (d) are horizontal coordinates of the peaks.}
\label{FigureT2}
\end{figure}

Numerical results on the sphere are presented in Fig.~\ref{FigureS2} and those on the torus in Fig.~\ref{FigureT2}. Energy spectra for $t=-0.45$ are displayed in panels (a). For the sphere, a branch of neutral excitations is identified at $L=2,3,4$. In the absence of tunneling, the Laughlin state of four bosons and the Moore-Read state of eight bosons both have neutral excitations at $L=2,3,4$. As in the previous example, tunneling merges them into a single branch. For the torus, we only show half of the momentum sectors because they are repeated once again in the other sectors due to a center of mass translation symmetry~\cite{Haldane1985b}. The total number of quasi-degenerate ground states is four as what is expected for the Read-Rezayi $\mathbb{Z}_{3}$ state. It can be seen from the ground state energy derivatives that a third-order transition occurs at $t_{c} \approx 0.20$. In the vicinity of the critical point, the von Neumann entanglement entropy can be fitted using the scaling function in Eq.~\eqref{EntropyScaleFunc} with $\alpha=0.974$.

\subsection{Trilayer systems with $\nu_{\mathsf{t}}=1/2$, $\nu_{\mathsf{m}}=1/2$, and $\nu_{\mathsf{b}}=1/2$}

\begin{figure}[ht]
\centering
\includegraphics[width=0.75\textwidth]{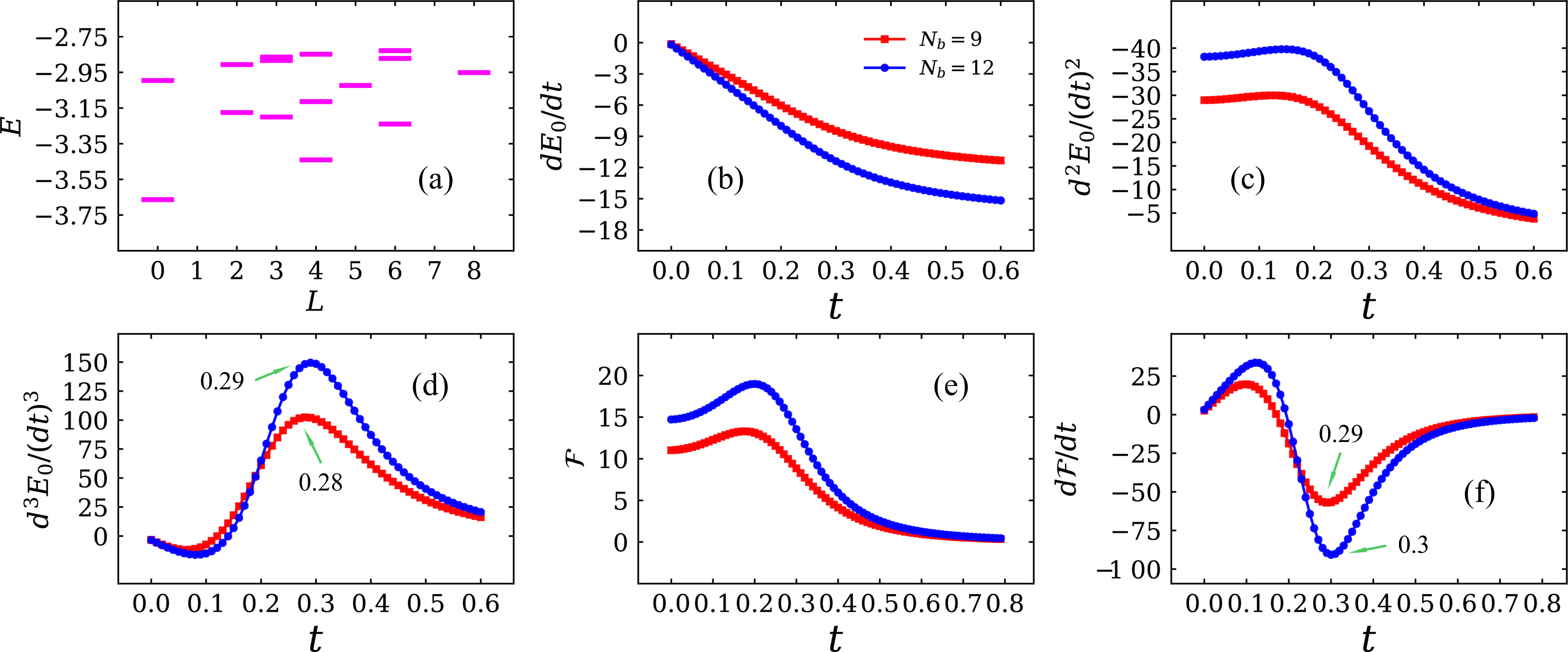}
\caption{Numerical results about trilayer systems with $\nu_{\mathsf{t}}=1/2,\nu_{\mathsf{m}}=1/2,\nu_{\mathsf{b}}=1/2$ on the sphere. (a) Energy spectrum for the case with $N_{b}=12$ and $t=-0.45$. (b-d) First-, second-, and third-order derivatives of the ground state energy. (e-f) Fidelity susceptibility of the ground state and its first-order derivative. The numbers in panels (d) and (f) are horizontal coordinates of the peaks.}
\label{FigureS3}
\end{figure}

\begin{figure}[ht]
\centering
\includegraphics[width=0.95\textwidth]{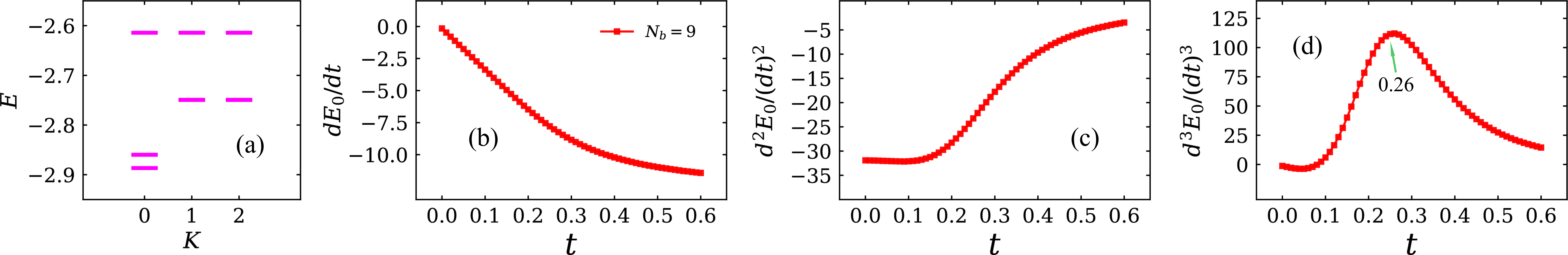}
\caption{Numerical results about trilayer systems with $\nu_{\mathsf{t}}=1/2,\nu_{\mathsf{m}}=1/2,\nu_{\mathsf{b}}=1/2$ on the torus. (a) Energy spectrum for the case with $N_{b}=9$ and $t=-0.45$. (b-d) First-, second-, and third-order derivatives of the ground state energy. The numbers in panel (d) are horizontal coordinates of the peaks.}
\label{FigureT3}
\end{figure}

In this example, the system realizes three decoupled Laughlin states in the absence of interlayer tunneling and turns into a Read-Rezayi $\mathbb{Z}_{3}$ state when the tunneling becomes sufficiently large. We study the many-body Hamiltonian
\begin{eqnarray}
H_{3} = V_{0}(\mathsf{t}) + V_{0}(\mathsf{m}) + V_{0}(\mathsf{b}) - t \sum_{k} \left[ C^{\dag}_{k}(\mathsf{t}) C_{k}(\mathsf{m}) + C^{\dag}_{k}(\mathsf{m}) C_{k}(\mathsf{t}) + C^{\dag}_{k}(\mathsf{m}) C_{k}(\mathsf{b}) + C^{\dag}_{k}(\mathsf{b}) C_{k}(\mathsf{m})  \right].
\end{eqnarray}
As in the first example, only zeroth Haldane pseudopotentials are used in each layer. There are tunneling terms between the top and middle layers as well as between middle and bottom layers but no direct tunneling connects the top and bottom layers. Numerical results on the sphere are presented in Fig.~\ref{FigureS3} and those on the torus in Fig.~\ref{FigureT3}. The findings are very similar to those of the previous examples: a third-order phase transition is unveiled from the ground state energy derivatives and the fidelity susceptibility derivative. On the large $t$ side, neutral excitations on the sphere can be obtained from those of the Laughlin states and there are four quasi-degenerate ground states on the torus. We note that this transition is fine tuned in some sense. If the two tunneling strengths are very different, one would expect that a Moore-Read state appears first in the two strongly coupled layers as the strength increases. In this case, there is no direct transition between the decoupled Laughlin states and the Read-Rezayi $\mathbb{Z}_{3}$ state.

\subsection{Chern-Simons-Higgs theory}

Since universal properties of phase transitions are generally captured by quantum field theory, it is useful to construct field theories for our systems. For all FQH states discussed above, the low-energy physics is described by Chern-Simons theory. The Laughlin state is usually described by the U(1)$_{2}$ Abelian Chern-Simons theory~\cite{WenXG-Book}, but the SU(2)$_{1}$ non-Abelian Chern-Simons theory is more convenient for our purpose~\cite{WenXG1998}. More generally, the Lagrangian density of SU(2)$_{k}$ non-Abelian Chern-Simons theory is
\begin{eqnarray}
\mathcal{L}_{k} = \frac{k}{4\pi} {\rm Tr} \left[ a{\wedge}da + \frac{2}{3} a{\wedge}a{\wedge}a \right],
\end{eqnarray}
where the gauge field $a$ takes values in the su(2) Lie algebra~\cite{Fradkin1998,Fradkin2001,Nayak2008}. The Laughlin, Moore-Read, and Read-Rezayi $\mathbb{Z}_{3}$ states correspond to $k=1$, $k=2$, and $k{\geq}3$, respectively. 

In the two bilayer examples, the top layer has Chern-Simons level $k_{\mathsf{t}}$ and the bottom layer has $k_{\mathsf{b}}$. We may construct a non-Abelian Chern-Simons-Higgs theory
\begin{eqnarray}
\frac{k_{\mathsf{t}}}{4\pi} {\rm Tr} \left[ a_{\mathsf{t}}{\wedge}da_{\mathsf{t}} + \frac{2}{3} a_{\mathsf{t}}{\wedge}a_{\mathsf{t}}{\wedge}a_{\mathsf{t}} \right] + \frac{k_{\mathsf{b}}}{4\pi} {\rm Tr} \left[ a_{\mathsf{b}}{\wedge}da_{\mathsf{b}} + \frac{2}{3} a_{\mathsf{b}}{\wedge}a_{\mathsf{b}}{\wedge}a_{\mathsf{b}} \right] + |\left(\partial-ia_{\mathsf{t}}+ia_{\mathsf{b}}\right)\phi|^{2} + r|\phi|^{2} + u|\phi|^{4}
\end{eqnarray}
with $\phi$ being a complex scalar field and $u>0$. The Higgs boson is gapped when $r>0$ but condenses when $r<0$. In the latter case, the gauge fields $a_{\mathsf{t}}$ and $a_{\mathsf{b}}$ can be set as equal to generate a SU(2)$_{k_{\mathsf{t}}+k_{\mathsf{b}}}$ theory. In the trilayer example, we can introduce two Higgs bosons that coupled respectively to $a_{\mathsf{t}}-a_{\mathsf{m}}$ and $a_{\mathsf{m}}-a_{\mathsf{b}}$. It is essential that they condense simultaneously so all gauge fields are set as equal.

\section{Other Phases}
\label{other}

If we also introduce interlayer interactions (its strength is characterized by $u$), the phase diagrams would become even richer. When interlayer interaction is strong but interlayer tunneling does not exist, the physics can be analyzed using the composite fermion theory. For the bilayer system at $\nu_{\mathsf{t}}=1/2,\nu_{\mathsf{b}}=1/2$, a two-component composite fermion liquid (CFL) has been found~\cite{WuYH2015b,Geraedts2017}. For the bilayer systems at $\nu_{\mathsf{t}}=1/2,\nu_{\mathsf{b}}=1$, a two-component Jain state has been found~\cite{WuYH2013}. For the trilayer system $\nu_{\mathsf{t}}=1/2,\nu_{\mathsf{m}}=1/2,\nu_{\mathsf{b}}=1/2$, a three-component Jain state could appear. The last state was not studied before, but we have confirmed its existence by numerical calculations. In all states, each boson is dressed by one flux to become composite fermions that experience an effective magnetic field. For the CFL, this field is zero so the composite fermions form a Fermi sea. For the other two cases, the composite fermions have effective filling factors $3$.

It is natural to explore the two-dimensional phase diagrams with interlayer interaction and tunneling as their two axes. Each one contains at least three states that may intersect at tricritical point. We have tried to map out the diagrams accurately but there are substantial challenges. The states discussed in Sec.~\ref{third} have shifts $2$ but the composite fermion states do not. If rotational symmetry is preserved (such as on the sphere or an infinite disk), there cannot be direct continuous transitions between states with different shifts~\cite{WenXG1992b}. Numerical calculations can still be done on the torus because it has only discrete rotational symmetry in finite-size systems. The results should be analyzed with care because full rotational symmetry is recovered in the thermodynamic limit. 

We have performed calculations along many line cuts in the $t-u$ space (fixing one variable and tuning the other) but the results are elusive. Let us explain the problem at $\nu_{\mathsf{t}}=1/2,\nu_{\mathsf{b}}=1/2$. Since the CFL is gapless, there are multiple low-energy eigenstates. The lowest one may not always stay in the same momentum sector and sometimes low-energy levels cross with each other. It is difficult to ascertain the onset of CFL. Along the lines of fixed $t$ and varying $u$, the derivatives of the lowest eigenvalue only have very weak features (shallow peaks), whose positions change considerably with $t$. Along the lines of fixed $u$ and varying $t$, we find clear signals of third-order phase transitions at small $u$, and the critical values of $t$ are close to $0.30$. For sufficiently large $u$, it seems that two transitions occur as $t$ increases. This may be interpreted as the system starts in the CFL phase, enters an intermediate phase, and eventually realize the Moore-Read state. We do not understand the nature of this intermediate phase and could not determine if there is a multi-critical point.

\section{Conclusions}
\label{conclusion}

To sum up, we have studied three third-order quantum phase transitions of bosonic non-Abelian FQH states in multi-layer systems. There are not many known examples of third-order transitions and the fact that our systems are strongly correlated makes their appearance quite surprising. In the absence of order parameters, we inspect the scaling behavior of entanglement entropy using purely empirical ansatz. Analytical support is highly desirable but difficult to achieve given the strongly correlated nature of the Chern-Simons-Higgs theory. As a byproduct, we speculate that certain modifications of fidelity susceptibility may be needed to probe higher-order phase transitions. It would be interesting to check if similar physics also appear in lattice models. This could provide a useful route toward realizing non-Abelian states and explore the critical properties in greater detail. 

An important question about the transitions is whether conformal symmetry emerges at the critical points such that the systems are described by (2+1)D conformal field theory (CFT). In the past few years, it has been found that certain (2+1)D CFTs can be realized as quantum phase transitions in the LLs on the sphere~\cite{ZhuW2023}. This is very helpful for conducting accurate quantitative investigations. For the $\nu_{\mathsf{t}}=1/2,\nu_{\mathsf{b}}=1/2$ case, this has been done in Ref.~\cite{Voinea2025b}. The other two cases should also be scrutinized in this regard.

\vspace{1em}

{\it Note added} --- During the preparation of this manuscript, we became aware of a preprint that also studies tunneling driven phase transition~\cite{Voinea2025b}. It focuses on the $\nu_{\mathsf{t}}=1/2,\nu_{\mathsf{b}}=1/2$ case and revealed the presence of conformal symmetry at the critical point. We learned from this work that a gauged 3D Majorana fermion theory was proposed for the transition~\cite{WenXG2000,Barkeshli2011}. The connection with our Chern-Simons-Higgs theory is unclear at present.

\section*{Acknowledgments}

We thank Meng Cheng, Guo-Zhu Liu, Hong-Hao Tu, and Wei Zhu for helpful discussions. Some calculations are performed using the DiagHam package for which we are grateful to its authors~\cite{DiagHam}. This work was supported by the NNSF of China under grant No.~12174130.

\bibliography{../ReferConde}

\end{document}